\documentclass[a4paper, smallextended]{svjour3}
\usepackage{bbding}
\usepackage[T1]{fontenc}
\usepackage{aurical}
\usepackage{huncial}
\usepackage{linearb}
\usepackage{textcomp}
\usepackage{amsmath}
\usepackage{amssymb}
\usepackage{amsfonts}
\usepackage{amssymb,epic,eepic}
\usepackage{stmaryrd}
\usepackage{capt-of}
\usepackage{graphicx}
\usepackage[usenames,dvipsnames]{pstricks}
\usepackage{epsfig}
\usepackage{pst-grad}
 \usepackage{pst-plot}
\usepackage{here}
\usepackage[justification=centering]{caption}
\smartqed

\begin{document}
\title{The broadcast classical-quantum capacity region 
of a two-phase
bidirectional relaying channel}
\author{Holger Boche \and Minglai Cai \and Christian Deppe}
\institute{Lehrstuhl f\"ur Theoretische
Informationstechnik,\\ Technische Universit\"at M\"unchen,\\
Munich, Germany\\
\email{\{boche, minglai.cai, christian.deppe\}@tum.de}}
\titlerunning{Bidirectional Relaying Classical-Quantum Channel}
\maketitle
\begin{abstract} We studied
a three-node quantum network that  enables
bidirectional communication between
two nodes with a half-duplex relay node for transmitting classical messages.
A decode-and-forward protocol is used to perform the communication in two phases.
In the first phase, the messages of two
nodes are transmitted to the relay node. The capacity of the first 
 phase is well-known by previous works. In the second phase, the relay node
broadcasts a re-encoded composition to the two
nodes. 
We determine the capacity 
region of the broadcast phase. To the best of our knowledge, this is the first paper analyzing quantum
bidirectional relay networks. 
\end{abstract}

\section{Introduction}\label{tsoqcnhbmamiitlfysofao}
The study  of quantum channel networks has become more and more 
important in the last few years. 

Some of the first applications of this will be
secret key transmission/gene\-ration and transmitting of
secure messages over quantum networks. 
The capacities for secret key transmission/gene\-ration and the
secrecy capacities for message transmission
have been
determined in \cite{De} and \cite{Ca/Wi/Ye}.
 
For these applications, the transmitters have to 
solve two main problems. First, the message (a secret key
or a secure message)
has to be encoded in such a way that
it
 can be decoded correctly by the legal receiver.
Second, the message has to be encoded such that
the wiretapper's knowledge of the 
transmitted classical message can be 
kept arbitrarily small.

Transmission of secret keys
and  secure messages
 over long distances is an essential requirement
for those applications.
Thus, in our paper, we consider  the problem how to ensure that
the legal receivers are able to reproduce
the original messages. This is a necessary condition for
reliable secret key transmissions/gene\-rations and transmitting of
secure messages over quantum networks.

The problem of long-distance transmissions of quantum information is one of 
the biggest problems in the realization of quantum networks.
The  sending of photons in optical fibers
is presently limited by 200 km because of losses in the optical-fiber channel due to 
absorption on the way. One solution to solve this problem is the
development of quantum repeaters (cf. \cite{Br/Due/Ci/Zo} and
\cite{Ab/Br/Be/Ka/vLo/Br}). Unfortunately, the 
practical realization of this component is not given until now.
The researchers already built quantum repeaters in their laboratories, but 
until now it is not possible to extend the limit of 200 km.

In this paper, we use a relay instead of a quantum repeater.
 The advantage 
of this protocol is that it is realizable and enables quantum 
communication
between two parties over the double length of the distance for
sending photons in optical fibers. This protocol
can also be used for free space optical communications
between satellites.
The communications in several classical practical applications such as
satellite communication and cellular communication are
modeled with channels with relay nodes. 
Channel networks with relay nodes have been studied extensively in the
context of classical information theory (cf. \cite{Kr/Ga/Gu} and \cite{Co/Ga}).
The study of quantum channels with relay nodes has just
recently begun  (cf. \cite{Sa/Wi/Vu}).

\begin{center}\begin{figure}[H]\centering
\includegraphics[width=0.4\linewidth]{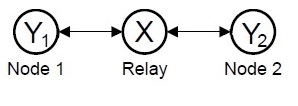}
\caption{
Two-phase bidirectional relaying channels}\label{bidirelpic}\end{figure}\end{center}

We analyze a
 quantum channel network model which was introduced  for classical channel networks
in \cite{Oech/Sch/Bj/Bo}. 
It is called the two-phase bidirectional relaying channel
(Figure \ref{bidirelpic}).
 In this model we consider a
three-node quantum network with two message sets $M_1$ and $M_2$,
which is called a two-user bidirectional quantum channel. The message
$m_2 \in M_2$ is located at node~$1$, and the message $m_1
\in M_1$ is 
located at node~$2$, respectively,
while a relay node  enables the bidirectional
communication between these nodes.
We assume that the relay node cannot transmit and receive data at the same time.
This is usually called a half-duplex relay. This assumption
is reasonable for practical components in 
communication systems in general.

Our goal is that after the transmission
the message $m_2 \in M_2$ is known at node~$2$
 and the message $m_1
\in M_1$  is known at node~$1$,
respectively. We simplify the problem by assuming 
an a priori
separation of the communication into two phases.

There exist several strategies which are usually classified
by the process at the relay node, namely the entanglement 
swapping-and-forward strategy (cf. \cite{Ja/Pi/Fr}) and the decode-and-forward strategy. 
We
consider a  two-phase decode-and-forward
protocol. The relay
node's task is to  decode the messages that it receives in the first phase
and to forward the information to its destinations in the second phase.
This basically means that in the first phase the relay measures and decodes and 
that in the second phase the relay prepares and encodes. The disadvantage of this
strategy is that the coherence is destroyed. 
The advantage of  this strategy
is that in the second phase
each receiver can
use its own message from the first phase as side information
to gather a higher capacity.

The goal
of the communication is 
the transmission of
classical signals, between two partners.  
The problem of
 the optimal transmission of
these classical signals
can be divided
into two  parts:
 \begin{enumerate}
\item Quantum modulation (i.e.,
choosing an 
 optimal set
of possible input states which will
be the input alphabet. This 
is equivalent  to  consider a special classical-quantum channel,
i.e., a quantum  channel
depending on the set of chosen input states
whose sender's inputs are classical variables.)
\item  Optimal coding for the classical-quantum channel.
   \end{enumerate}

We also consider this problem
because  the model of classical-quantum channels
is a very important tool for understanding
the capacity formulas for quantum channels. 
The capacity
of  classical-quantum channels has been  determined   in \cite{Ho}, \cite{Ho2},
and \cite{Sch/We}.
It turns out that classical-quantum channels are not only of theoretical interest,
but as well as of technical interest, too; for instance, a code
for a classical-quantum channel can be used for entanglement generation (cf. \cite{De}).
Furthermore, new phenomenons such as super-additivity and super-activation appear for 
carrying classical information through a quantum channel (cf. \cite{Bo/Ca/De}).

We would like to point out that the three-node bidirectional relay network
is
 an extremely
advantageous tool in the relay network theory.
The
model of classical three-node bidirectional relay network
has  been extended to more complex models such as
MIMO channels (\cite{Oech/Jo/Wy/Bo}, \cite{Do/Oech/Sk}, \cite{Wy/Oe/Bo}, \cite{Oech/Wy/Bo}),
Gaussian channels (\cite{Wy/Oe/Bo}), 
polar codes (\cite{An/Wy/Oe/Sk}), and cross-layer designs (\cite{Oech/Bo}, \cite{Oech/Bo2}, \cite{Oech/Bo3}).
In view of these previous works 
on the classical bidirectional relay network,
our further  tasks
will be to analyze these models
for quantum networks (cf.
\cite{Is/Hi}, \cite{Ei/Wo}, and \cite{Wil/Gu}).

For classical models, the authors of \cite{Oech/Sch/Bj/Bo}
use results from coding theory for degraded broadcast channel (cf. \cite{Ber}).
One of our major challenges in this paper is that there are no  equivalent tools
in the quantum information theory yet. Thus, 
we can only establish
 the capacity region
 with average errors, but not the capacity region
with maximal errors as in the classical case 
(cf. Remark \ref{avemax}). It still reminds open how
 the capacity region
with maximal errors can be established.

Recent research has been done toward
security
for classical bidirectional   relay networks
(cf. \cite{Wy/Wi/Bo} and \cite{An/Wy/Oe/Sk}).
It is a promising task
to find similar results for quantum networks.

Another important  basic feature for classical bidirectional
 relay networks is the
channel uncertainty (\cite{Wy/Bj/Oe/Bo}, \cite{Oech/Sk}).
The capacity for quantum channels with uncertainty
has been determined in
\cite{Ahl/Bli}, \cite{Bj/Bo/Ja/No}, and \cite{Bo/Ca/De}.

The design of communication protocols for quantum networks
is a challenging task. For the design of efficient protocols,
it is important to incorporate side information into the coding
schemes. In this paper,  we considered the case where both
receivers have side information about the messages. 
\cite{Bo/Ca/No} considered recently the case where the 
transmitter has side information about the channel state.
This corresponds to the classical ``writing on dirty
paper'' coding. This coding strategy plays an important
role in modern communication sy\-stems. It is an interesting
research topic to develop on the basic of \cite{Bo/Ca/No}
and the  bidirectional relaying protocol of this paper
for more complicated quantum networks.

The study of relay channels in the quantum scenario is novel. 
We hope our results may raise interests in further analysis.


\section{Basic definitions}

\subsection{A two-phase protocol}

By the separation of the communication, we have a multiple-access
phase, where node~$1$ and node~$2$ transmit messages $m_2$ and $m_1$ to
the relay node, and a broadcast phase, where the relay forwards the
messages to node~$2$ and  node~$1$, respectively. We look at the two
phases separately.

\begin{center}\begin{figure}[H]\centering
\includegraphics[width=0.4\linewidth]{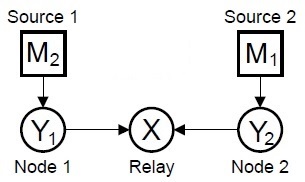}
\caption{
The multiple access phase}
\end{figure}\end{center}

In the multiple-access phase,  we have a classical-quantum multiple-access 
channel. The multiple-access channel is a channel such that
two (or more) senders send information to a common receiver via 
this channel. The
optimal coding strategies and capacity regions for   classical 
multiple-access channels have been given in
 \cite{Ahl4} and \cite{Lia}. 
The optimal coding strategies and capacity regions for
   multiple-access quantum channels have been given in
\cite{Win2} and  \cite{Ya/Ha/De2}.

\begin{center}\begin{figure}[H]\centering
\includegraphics[width=0.4\linewidth]{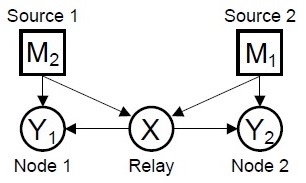}
\caption{
The broadcast phase}
\end{figure}\end{center}

In the broadcast phase,  we have a broadcast quantum
channel. In a broadcast channel, 
one single
sender sends 
information to 
two
(or more) receivers.  
The optimal
coding strategies and capacity regions for   classical broadcast channels
have been given in \cite{Ma}, \cite{Ber}, and \cite{Ga}. 
An optimal coding
strategy and a capacity region for broadcast quantum channels
have been given in \cite{Sa/Wi}.

 For the broadcast phase, we assume that the relay node has
successfully decoded the messages $m_1$ and $m_2$ in the multiple-access phase. Of course,
the message $m_2$ 
is also known at
node~$1$   
and the message $m_2$ is 
also known at node~$2$.

The goal of the relay node is to broadcast a message to node~$1$
and  node~$2$ which allows both nodes to recover the unknown source. This means
that node~$1$ wants to recover message $m_1$ and that node~$2$ wants to
recover message $m_2$.

\subsection{Notations and communication
scenarios}

 For  finite-dimensional
complex Hilbert spaces  $G$ and  $G'$,  a quantum channel $N$:
$\mathcal{S}(G) \rightarrow \mathcal{S}(G')$, $\mathcal{S}(G)  \ni
\rho \rightarrow N(\rho) \in \mathcal{S}(G')$ is represented by a
completely positive trace-preserving map
 that accepts input quantum states in $\mathcal{S}(G)$ and produces output quantum
states in  $\mathcal{S}(G')$. Here, $\mathcal{S}(G)$  stands for the space of  density operators on the
space $G$.\vspace{0.15cm}

If the sender wants to transmit a classical message of a finite set $A$ to
the receiver using a quantum channel $N$, his encoding procedure will
include a classical-to-quantum encoder 
to prepare a quantum message state $\rho \in
\mathcal{S}(G)$ suitable as an input for the channel. If the sender's
encoding is restricted to transmit an  indexed finite set of
 quantum states $\{\rho_{x}: x\in A\}\subset
\mathcal{S}(G)$, then we can consider the choice of the signal
quantum states $\rho_{x}$ as a component of the channel. Thus, we
obtain a channel $\sigma_x := N(\rho_{x})$ with classical inputs $x\in A$ and quantum outputs,
 which we call a classical-quantum
channel. This is a map $\mathbf{N}$: $A \rightarrow
\mathcal{S}(G')$, $A \ni x \rightarrow \mathbf{N}(x) \in
\mathcal{S}(G')$ which is represented by the set of $|A|$ possible
output quantum states $\left\{\sigma_x = \mathbf{N}(x) :=
N(\rho_{x}): x\in A\right\}\subset \mathcal{S}(G')$, meaning that
each classical input of $x\in A$ leads to a distinct quantum output
$\sigma_x \in \mathcal{S}(G')$. In view of this, we have the following
definition.\vspace{0.15cm}

Let $H$ be a finite-dimensional
complex Hilbert space. A classical-quantum channel is a map $N:
A\rightarrow\mathcal{S}(H)$, $A  \ni a \rightarrow N(a) \in
\mathcal{S}(H)$.\vspace{0.15cm}

For a probability distribution $P$ on a finite set $A$  and a positive constant $\delta$,
we denote the set of typical sequences by 
\[\mathcal{T}^n_{P,\delta} :=\left\{ x^n \in A^n: \left\vert \frac{1}{n} \left\langle x^n \mid a\right\rangle - P(a) \right\vert \leq \frac{\delta}{|A|}\right\}\text{ ,}\]
where $\left\langle x^n \mid a\right\rangle$ is the number of occurrences of the symbol $a$ in the sequence $x^n$.\vspace{0.15cm}

Let  $n\in\mathbb{N}$, 
we define $A^n:= \{(a_1,\ldots,a_n): a_i \in A
\text{ } \forall i \in \{1,\ldots,n\}\}$. 
 The space which the vectors
$\{v_1\otimes \ldots \otimes v_n: v_i \in H
\text{ } \forall i \in \{1,\ldots,n\}\}$ span is defined 
by $H^{\otimes n}$. We also write $a^n$ 
for the elements of
$A^n$.

Associated with a classical quantum channel, $N$:
$A \rightarrow \mathcal{S}(H)$ is the channel map on the $n$-block $N^{\otimes n}$:
$A^n \rightarrow \mathcal{S}({H}^{\otimes n})$ such that for $a^n =
(a_1,\ldots,a_n) \in A^{ n}$. We have
 $N^{\otimes n}(a^n) = N(a_1)
\otimes \ldots \otimes N(a_n)$.\vspace{0.15cm}

For a quantum state $\rho\in \mathcal{S}(G)$, we denote the von Neumann
entropy of $\rho$ by \[S(\rho)=- \mathrm{tr}(\rho\log\rho)\text{
.}\] Let
$\mathbf{V}$: $A \rightarrow
\mathcal{S}(G)$ be a classical-quantum
channel. For $P\in P(A)$
the conditional entropy of the channel for $\mathbf{V}$ with input distribution $P$
is denoted by
 \[S(\mathbf{V}|P) := \sum_{x\in A} P(x)S(\mathbf{V}(x))\text{
.}\]
\begin{remark}The following definition is a more general definition of
the conditional entropy in  quantum information theory.
Let $\mathfrak{P}$ and $\mathfrak{Q}$ be quantum systems. We 
denote the Hilbert space of $\mathfrak{P}$ and $\mathfrak{Q}$ by 
$G^\mathfrak{P}$ and $G^\mathfrak{Q}$, respectively. Let $\phi^\mathfrak{PQ}$ be a bipartite
quantum state in $\mathcal{S}(G^\mathfrak{PQ})$. 
We denote
$S(\mathfrak{P}\mid\mathfrak{Q})_{\rho}:=
S(\phi^\mathfrak{PQ})-S(\phi^\mathfrak{Q})$.
Here $\phi^\mathfrak{Q}=\mathrm{tr}_{\mathfrak{P}}(\phi^\mathfrak{PQ})$.
\end{remark}\vspace{0.2cm}

Let 
$\varPhi := \{\rho_x : x\in A\}$  be a set of quantum  states labeled by
elements of $A$. For a probability distribution $P$ on $A$
the    Holevo $\chi$ quantity is
defined as
\[\chi(P;\varPhi):= S\left(\sum_{x\in A} P(X)\rho_x\right)-
\sum_{x\in A} P(X)S\left(\rho_x\right)\text{ .}\]

We denote the identity operator on a space $G$ by $\mathrm{id}_G$.

A collection of positive semi-definite operators $\{M_i:i\}$ on
${G}$ is called a positive operator valued measure,
or POVM, if it is a  partition of the identity, i.e., $\sum_{i} M_i
= \mathrm{id}_{G}.$\vspace{0.25cm}

\subsection{Code concepts}

A two-user multiple-access quantum channel  $N_{BC-A}:
H^{BC}\rightarrow H^A$ has two senders $B$ and $C$, and a single 
receiver $A$. It is defined  as a map $N:  H^{BC}\rightarrow
H^A$.

An $(n, J_n^{(1)},J_n^{(2)})$ code carrying classical information
for a  two-user quantum   multiple access channel $N_{BC-A}: H^{BC}\rightarrow
H^A$ consists of a  ensemble of quantum
states $\{w(m_1): m_1=1,\ldots,J_n^{(1)}\} \subset \mathcal{S}({H^{B}}^{\otimes n})$,
quantum states $\{v(m_2): m_2=1,\ldots,J_n^{(2)}\} \subset \mathcal{S}({H^{C}}^{\otimes n})$,
and a POVM $\biggl\{D_{m_1,m_2}: m_1\in \{ 1,\ldots ,J_n^{(1)}\}, m_2\in \{ 1,\ldots ,J_n^{(2)}\}\biggr\}$
on ${H^{A}}^{\otimes n}$.


A pair of nonnegative numbers $(R_1,R_2)$ is an achievable  rate pair  with
classical inputs for the  quantum   multiple-access  channel
$N_{BC-A}: H^{BC}\rightarrow
H^A$ with average
error if  for every positive $\varepsilon$,  $\delta$,
and a  sufficiently large $n$ there is  an  $(n, J_n^{(1)},J_n^{(2)})$
 code  carrying classical information
$(\{w(m_1):m_1=1,\ldots,J_n^{(1)}\},$ $\{v(m_2):m_2=1,\ldots,J_n^{(2)}\},\{D_{m_1,m_2}:
m_1\in \{ 1,\ldots ,J_n^{(1)}\},$ $m_2\in \{ 1,\ldots ,J_n^{(2)}\}\})$ such that
$\frac{1}{n}\log J_n^{(1)}$ $\geq R_1-\delta$,
$\frac{1}{n}\log J_n^{(2)}$ $\geq R_2-\delta$ and
\begin{equation}
 \frac{1}{J_n^{(1)}J_n^{(2)}} \sum_{m_1=1}^{J_n^{(1)}} \sum_{m_2=1}^{J_n^{(2)}}\mathrm{tr}\left((\mathrm{id}_{{H^{A}}^{\otimes n}}-D_{m_1,m_2})
N_{BC-A}^{\otimes n}\left((w(m_1),v(m_2))\right)\right)\leq \varepsilon\text{
.}\end{equation}

The two-user broadcast quantum channel $N_{A-BC}: H^A\rightarrow
H^{BC}$ is a quantum channel from a
 single  sender $A$
to two independent receivers $B$ and $C$. The quantum channel $W_1$ from
$A$ to $B$ is obtained by tracing out $C$ from the channel map,
i.e., $W_1=N_{A-B}:H^A\rightarrow H^{B}$, which is the quantum channel from $A$
to $B$, is defined as $W_1(\sigma) =\mathrm{tr}_{C}(N_{A-BC}(\sigma))$. Furthermore, $W_2=N_{A-C}: H^A\rightarrow
H^{B}$, which is the quantum channel from $A$ to $C$, is defined as $W_2(\sigma) =\mathrm{tr}_{B}(N_{A-BC}(\sigma))$.

An $(n, J_n^{(1)},J_n^{(2)})$ code carrying classical information
for a two-user broadcast quantum channel $N_{A-BC}$: $H^A$ $\rightarrow$
$H^{BC}$ consists of a  ensemble $\{w((m_1,m_2)): m_1=1,\ldots,J_n^{(1)},m_2=1,\ldots,J_n^{(2)}\}$ $\subset$ $\mathcal{S}({A}^{\otimes n})$,
 a POVM $\left\{D_{m_1}^{(1)}: m_1\in \{ 1,\ldots ,J_n^{(1)}\}\right\}$
on ${H^{B}}^{\otimes n}$,
and a POVM $\left\{D_{m_2}^{(2)}: m_2\in \{ 1,\ldots ,J_n^{(2)}\}\right\}$
on ${H^{C}}^{\otimes n}$.\vspace{0.15cm}


A pair of nonnegative numbers $(R_1,R_2)$ is an achievable  rate pair  with
a classical input for the  two-user broadcast quantum channel
$N_{A-BC}: H^A\rightarrow
H^{BC}$ with average
error if  for every positive $\varepsilon$,  $\delta$,
and a  sufficiently large $n$ there is  an  $(n, J_n^{(1)},J_n^{(2)})$
 code  carrying classical information
$(\{w_t(j):j\},\{D_j:j\})$ such that
$\frac{1}{n}\log J_n^{(1)} \geq R_1-\delta$,
$\frac{1}{n}\log J_n^{(2)} \geq R_2-\delta$, for every $m_2\in M_2$
\begin{equation}
 \frac{1}{J_n^{(1)}} \sum_{j=1}^{J_n^{(1)}} \mathrm{tr}\left((\mathrm{id}_{{H^{B}}^{\otimes n}}-D_{m_1}^{(1)})
W_1^{\otimes n}\left( w((m_1,m_2))\right)\right)\leq \varepsilon\text{
,}\end{equation} and for every $m_1\in M_1$
\begin{equation}
 \frac{1}{J_n^{(2)}} \sum_{j=1}^{J_n^{(2)}} \mathrm{tr}\left((\mathrm{id}_{{H^{C}}^{\otimes n}}-D_{m_2}^{(1)})
W_2^{\otimes n}\left( w((m_1,m_2))\right)\right)\leq \varepsilon\text{
,}\end{equation}
where $W_1=N_{A-B}$ and $W_2=N_{A-C}$.

The capacity regions of multiple-access quantum channels and broadcast
quantum channels are
 convex by the time-sharing principle: Let
$(R_1,R_2)$ and $(R_1',R_2')$ be rate tuples of $m$ and $n$ block
codes, respectively, with error probabilities $\epsilon_1$ and 
$\epsilon_2$, respectively. We
get an $(m + n)$ block code with error probability at most
$\epsilon_1+\epsilon_2$ and with rates
$\left(\frac{m}{m+n}R_1+\frac{n}{m+n}R_1',\frac{m}{m+n}R_2+\frac{n}{m+n}R_2'\right)$
 by concatenating the code words to $(m +
n)$ blocks and tensoring the corresponding decoding observables.

\section{The classical-quantum capacity region of 
the bidirectional relaying quantum channel}
For the multiple-access phase, the optimal coding strategy is well
known from \cite{Win2}, where the following lemma for the classical-quantum  rate
region of  multiple-access quantum channels was given.
\vspace{0.15cm}


\begin{lemma}
Let $N_{Y_2Y_1-X}$ be a two-user   multiple-access quantum channel. Let $H^{Y_1}$ be the
Hilbert space whose unit vectors correspond to the pure states of node~$1$'s quantum 
system,  $H^{Y_2}$ be the  
    Hilbert space whose unit vectors correspond to the pure states of node~$2$'s quantum
system, and $H^{X}$ be the
Hilbert space whose unit vectors correspond to the pure states of the relay node's quantum
system.

We assume   node~$1$'s encoding is restricted to transmitting an indexed
finite set of orthogonal quantum  states $Y_1\subset H^{Y_1}$.

We assume  node~$2$'s encoding is restricted to transmitting an indexed
finite set of orthogonal quantum  states $Y_2\subset H^{Y_2}$.

The classical-quantum capacity region  of the multiple-access quantum 
channel $N_{Y_2Y_1-X}$ with average error is given by the set of all rate pairs
$(R_2, R_1)$,
satisfying
\begin{equation} R_2 \leq  \chi(Q_1;\sigma^{X}) \text{ ,}
\label{Brate2a}\end{equation}
\begin{equation} R_1 \leq \chi(Q_2;\sigma^{X}) \text{ ,}
\label{Brate1a}\end{equation} and
\begin{equation} R_2 + R_1 \leq  \chi(Q_{1,2};\sigma^{X})\label{Brate3a}\end{equation}
for any joint probability distribution $Q_{1,2}$ on $Y_1\times Y_2$.
Here, $Q_1$ is the marginal probability distribution of $Q_{1,2}$ on $Y_1$,
 $Q_2$ is the marginal probability distribution of $Q_{1,2}$ on $Y_2$, and
$\sigma^{X}$ is the resulting quantum state at the outcome of the relay node.
\label{macphase}\end{lemma}

Thus, if $n$ is sufficiently large and if for $M_1$ and $M_2$ it
holds

\[ |M_2| \leq  \lfloor 2^{n(\chi(Q_1;\sigma^{X}
)-\epsilon)} \rfloor \text{ ,}\]
\[|M_1| \leq \lfloor 2^{n(\chi(Q_2;\sigma^{X}) -\epsilon)} \rfloor \text{ ,}\]
 and \[|M_2| + |M_1| \leq \lfloor
2^{n(\chi(Q_{1,2};\sigma^{X})-\epsilon)} \rfloor \text{ }\]
for some positive $\epsilon$,
 we can
assume 
that
the relay node has successfully decoded
the messages $m_1 \in M_1$ and $m_2 \in M_2$.

Note that the author of \cite{Win2}, using block codes and showing
a weak converse,
is able to give 
(\ref{Brate2a}),  (\ref{Brate1a}), and (\ref{Brate3a})
 in single letter formula.
Please see \cite{Bo/Ca/No} for more discussions on the value 
of multiletter formulas for quantum communication networks.
\vspace{0.2cm}

For the broadcast phase, 
since each node has
perfect knowledge about the message intended for the other,
one can 
 use this knowledge as a support for the choice of the
decoding strategy to decode the message intended for itself.
In view of
these facts, we have the following Theorem
\ref{broadphase}.\vspace{0.15cm}

\begin{theorem}
Let $N$ be a two-user bidirectional quantum channel.
 Let $H^{Y_1}$ be the
Hilbert space whose unit vectors correspond to the pure states of node~$1$'s quantum
system,  $H^{Y_2}$ be the
Hilbert space whose unit vectors correspond to the pure states of node~$2$'s quantum
system, and $H^{X}$ be the
Hilbert space whose unit vectors correspond to the pure states of the relay node's quantum
system. Let $N_{X-Y_1Y_2}$ be the broadcast quantum channel in the broadcast phase.

We assume that  the relay node's encoding is restricted to transmitting
an indexed finite set of orthogonal quantum  states $\{\phi_x:
x\in X\} \subset H^X$.

For all probability
distribution $P$ on $X$,
the capacity region  of the bidirectional broadcast quantum channel $N_{X-Y_1Y_2}$
during the broadcast phase for transmitting classical information
with average error is given by the set of all rate pairs $(R_1, R_2)$, satisfying
\begin{equation} R_1 \leq  \limsup_{n\rightarrow \infty}\frac{1}{n} \chi(P^n;{\sigma^{Y_1}}^{\otimes n})   \text{ }
\label{Brate1b}\end{equation} and
\begin{equation} R_2 \leq   \limsup_{n\rightarrow \infty} \frac{1}{n}\chi(P^n;{\sigma^{Y_2}}^{\otimes n})  \text{ .}
\label{Brate2b}\end{equation}
Here, $\sigma^{Y_1}$ is the resulting quantum state at the outcome of node~$1$,
while $\sigma^{Y_2}$ is the resulting quantum state at the outcome of node~$2$.\label{broadphase}\end{theorem}

\proof
It is easy to verify that every achievable rate pair cannot exceed  
(\ref{Brate1b}) and  (\ref{Brate2b}). $R_1$ cannot exceed
$\limsup_{n\rightarrow \infty} \frac{1}{n} \chi(P^n;{\sigma^{Y_1}}^{\otimes n})$,
even if the relay node only sends a message to 
node~$1$ without sending any message to node $2$ (cf. \cite{Ho2}). For the same reason, $R_2$ cannot exceed
$\limsup_{n\rightarrow \infty} \frac{1}{n} \chi(P^n;{\sigma^{Y_2}}^{\otimes n})$ either.
Now we will prove
the achievability of the extremal point of the rate region given by
(\ref{Brate1b}) and  (\ref{Brate2b}), since then every rate pair in the rate region  is
achievable by the time-sharing principle.\vspace{0.2cm}

At first, we present some tools which   that were  used for our proof: \vspace{0.15cm}

Let $H$ be a  Hilbert space.
For $\rho \in \mathcal{S}(H)$ and $\alpha > 0$ there exists an
orthogonal subspace projector $\Pi_{\rho ,\alpha}$ commuting with
$\rho ^{\otimes n}$ and satisfying
\begin{equation} \label{te1} \mathrm{tr} \left( \rho^{\otimes n}
 \Pi_{\rho ,\alpha} \right) \geq 1-\frac{d}{4n\alpha^2}\text{ ,}\end{equation}
\begin{equation} \label{te2} \mathrm{tr} \left( \Pi_{\rho ,\alpha} \right)
 \leq 2^{n S(\rho) + Kd\alpha \sqrt{n}}\text{ ,}\end{equation}
\begin{equation} \label{te3}  \Pi_{\rho ,\alpha} \cdot \rho ^{\otimes n} \cdot \Pi_{\rho ,\alpha} \leq
2^{ -nS(\rho) + Kd\alpha \sqrt{n}}\Pi_{\rho ,\alpha}\text{
,}\end{equation} where  $d := \dim H$ and $K$ is a positive  constant (cf. \cite{Wil}).\\ 
Let $A$ be a finite set and let $\mathbf{V}: A \rightarrow \mathcal{S}(H)$ be a classical-quantum channel. 
For a probability
distribution $P$ on $A$, $\alpha > 0$, and $x^n \in \mathcal{T}^n_P$  there
exists an orthogonal subspace projector $\Pi_{\mathbf{V},\alpha}(x^n)$
commuting with $\mathbf{V}^{\otimes n}_{x^n}$ and satisfying
\begin{equation} \label{te4}  \mathrm{tr} \left( \mathbf{V}^{\otimes n}(x^n) \Pi_{\mathbf{V},\alpha}(x^n) \right)
 \geq 1-\frac{ad}{4n\alpha ^2}\text{ ,}\end{equation}
\begin{equation} \label{te5} \mathrm{tr} \left( \Pi_{\mathbf{V},\alpha}(x^n) \right)
\leq 2^{n S(\mathbf{V}|P) + Kad\alpha \sqrt{n}}\text { ,}\end{equation}
\begin{align}   &  \Pi_{\mathbf{V},\alpha}(x^n) \cdot \mathbf{V}^{\otimes n}(x^n) \cdot \Pi_{\mathbf{V},\alpha}(x^n) \notag\\
&\leq 2^{ -nS(\mathbf{V}|P) + Kad\alpha
\sqrt{n}}\Pi_{\mathbf{V},\alpha}(x^n)\text{ ,}\label{te6} \end{align}
where $a := \#\{A\}$  and $K$ is a positive constant (cf. \cite{Wil}).\\
Let $\mathbf{V}: A \rightarrow \mathcal{S}(H)$ be a classical-quantum channel.
 Then, every probability
distribution $P$ on $A$ defines a quantum state $P\mathbf{V}$ on $\mathcal{S}(H)$, which
 is the resulting quantum state at the output of $\mathbf{V}$ when the
input is sent according to $P$. Thus, for $\alpha' > 0$ we can define an 
orthogonal subspace projector $\Pi_{P\mathbf{V}, \alpha' \sqrt{a}}$
which fulfills (\ref{te1}), (\ref{te2}), and (\ref{te3}) (here, we set
$\rho= P\mathbf{V}$ and $\alpha = \alpha' \sqrt{a}$).
Furthermore, for $\Pi_{P\mathbf{V}, \alpha' \sqrt{a}}$, we have the
following inequality
\begin{equation} \label{te7}  \mathrm{tr} \left(  \mathbf{V}^{\otimes n}(x^n) \cdot \Pi_{P\mathbf{V}, \alpha' \sqrt{a}} \right)
 \geq 1-\frac{ad}{4n\alpha^2}\text{ ,}\end{equation}
where $K$ is a positive constant (cf. \cite{Wil}).
 \vspace{0.15cm}

\begin{lemma}[Measurement on Approximately Close States, cf. \cite{Wil}] \label{moacs} 
Let $\sigma$ and $\rho$ be two quantum states,  and let $\Pi$ be a
positive operator such that $\Pi \leq \mathrm{id}$; then,
\[\mathrm{tr}(\Pi\sigma) \geq \mathrm{tr}(\Pi\rho) - \|\sigma-\rho\|_1\text{ .}\]
\end{lemma}

\begin{lemma} [Tender Operator, cf. \cite{Win} and \cite{Og/Na}] \label{eQ_4a}
Let $\rho$ be a  quantum state. Let $X$ be a positive operator such that $X  \leq
\mathrm{id}$  and $1 - \mathrm{tr}(\rho X)  \leq
\lambda \leq 1$; then, 
\begin{equation} \| \rho -\sqrt{X}\rho \sqrt{X}\| \leq \sqrt{8\lambda}\text{ .}
\end{equation}\end{lemma}

\begin{lemma}[Hayashi-Nagaoka Operator Inequality, cf. \cite{Ha/Na}]\label{hnopin}
For any positive operators $S$ and $T$ such that $S \leq \mathrm{id}$ we have
\begin{equation} \mathrm{id}-(S + T)^{-\frac{1}{2}}S(S + T)^{-\frac{1}{2}} \leq (\mathrm{id} - S) + 4T\text{ .}
\end{equation}
\end{lemma}
\vspace{0.25cm}

\it The random encoding  technique: \rm\vspace{0.2cm}

We denote $W_1=N_{X-Y_1}$ and $W_2=N_{X-Y_2}$.
For any positive $\epsilon$
let ${M'}_1$ be a message set such that $|{M'}_1|\leq 2^{n(\chi(P;\sigma^{Y_1} )-2\epsilon)}$,
and let ${M'}_2$ be a message set such that $|{M'}_2|\leq 2^{n(\chi(P;\sigma^{Y_2} )-2\epsilon)}$.
We generate $|{M'}_1||{M'}_2|$ independent random variables
\[ \left\{X^n({m}_1, {m}_2): {m}_1\in {M'}_1, {m}_2\in {M'}_2\right\}\]
taking values in $\mathcal{T}^n_P$ i.i.d. according to the product distribution
$P(x^n) = \prod_{i=1}^{n} P(x_i)$.

For all $x^n$ $\in X^n$ we define $\Pi_{PW_1, \alpha \sqrt{a}}$ on ${H^{Y_1}}^{\otimes n}$,
$\Pi_{PW_2, \alpha \sqrt{a}}$ on ${H^{Y_2}}^{\otimes n}$, $\Pi_{W_1^{\otimes n}(x^n),\alpha}$ on ${H^{Y_1}}^{\otimes n}$,
and $\Pi_{W_2^{\otimes n}(x^n),\alpha}$ on ${H^{Y_2}}^{\otimes n}$ as in (\ref{te4}),
(\ref{te5}), (\ref{te6}), and (\ref{te7}). Here, we set $P=P$,
$\mathbf{V}=W_1$, and $=W_2$, respectively. $\alpha$ is some
positive constant which we will choose later. We define
\[{D'}^{(1)}_{x^n}:=\Pi_{PW_1, \alpha \sqrt{a}}\Pi_{W_1^{\otimes n}(x^n),\alpha}
\Pi_{PW_1, \alpha \sqrt{a}}\text{ ,}\] and
\[{D'}^{(2)}_{x^n}:=\Pi_{PW_2, \alpha \sqrt{a}}\Pi_{W_2^{\otimes n}(x^n),\alpha}
\Pi_{PW_1, \alpha \sqrt{a}}\text{ .}\] \vspace{0.2cm}

\it Analysis of errors of the first kind:\rm\vspace{0.2cm}

We say an error of the first kind occurs if $(m_1,m_2)$ has been send by
the relay node, and either node $1$ fails to decode $m_1$ or
node $2$ fails to decode $m_2$.\vspace{0.15cm}

For all $(m_1,m_2)\in {M'}_1\times {M'}_2$ and any realization
$x^n(m_1,m_2)$ of $X^n(m_1,m_2)$
we have

\begin{align}&\mathrm{tr}\left(W_1^{\otimes
n}(x^n(m_1,m_2)){D'}_{x^n(m_1,m_2)}^{(1)}\right)\allowdisplaybreaks\notag\\
&=\mathrm{tr}\Bigl( W_1^{\otimes
n}(x^n(m_1,m_2))
\Pi_{PW_1, \alpha \sqrt{a}}\Pi_{W_1^{\otimes n}(x^n(m_1,m_2)),\alpha}
\Pi_{PW_1, \alpha \sqrt{a}} \Bigr)\allowdisplaybreaks\notag\\
&=\mathrm{tr}\Bigl((\Pi_{PW_1, \alpha \sqrt{a}} W_1^{\otimes
n}(x^n(m_1,m_2))
  \Pi_{PW_1, \alpha \sqrt{a}})\Pi_{W_1^{\otimes n}(x^n(m_1,m_2)),\alpha}
 \Bigr)\allowdisplaybreaks\notag\\
&\geq \mathrm{tr}\Bigl( W_1^{\otimes
n}(x^n(m_1,m_2))\Pi_{W_1^{\otimes n}(x^n(m_1,m_2)),\alpha}
 \Bigr) \notag\\
&-\left\|\Pi_{PW_1, \alpha \sqrt{a}} W_1^{\otimes
n}(x^n(m_1,m_2))
  \Pi_{PW_1, \alpha \sqrt{a}}- W_1^{\otimes
n}(x^n(m_1,m_2))\right\|_1 \allowdisplaybreaks\notag\\
&\geq 1-\frac{d}{4n\alpha ^2}
\mathrm{tr}\Bigl(\Pi_{W^{\otimes n}(x^n(m_1,m_2)),\alpha} \Bigr) \notag\\
&-\left\|\Pi_{PW_1, \alpha \sqrt{a}} W_1^{\otimes
n}(x^n(m_1,m_2))
 \Pi_{PW_1, \alpha \sqrt{a}}- W_1^{\otimes
n}(x^n(m_1,m_2))\right\|_1 \allowdisplaybreaks\notag\\
&\geq 1-\frac{d}{4n\alpha ^2}
-\sqrt{8\frac{ad}{4n\alpha ^2}} \text{ .}\label{longineq2}
\end{align}
The first inequality holds because of Lemma \ref{moacs},
the second inequality holds  because of (\ref{te4}), and
the third inequality holds  because of Lemma \ref{eQ_4a}
and (\ref{te7}).\vspace{0.15cm}

Similarly, we have \begin{equation} \mathrm{tr}\left(W_2^{\otimes
n}(x^n(m_1,m_2)){D'}_{x^n(m_1,m_2)}^{(2)}\right) \geq 1-\frac{d}{4n\alpha
^2} -\sqrt{8\frac{ad}{4n\alpha ^2}}
 \text{ .}\label{longineq2b}\end{equation}
Thus, the errors of the first kind go to zero if $n$ is sufficiently large.
\vspace{0.2cm}

\it Analysis of errors of the second kind:\rm\vspace{0.2cm}

We define $\rho_2 := PW_2= \sum_{x\in X} P(X)W_2(\phi_x)$; then,
$\rho^{Y_2} = \rho_2$ if  any realization of $X^n$ is used to decode
the input message. Let us fix $(m_1,m_2)$, $(m_1,m_2')\in
{M'}_1\times {M'}_2$ such that $m_2\not= m_2'$. Node $2$ would make
an error if $(m_1,m_2)$ has been sent, but 
node $2$'s decoding results in the message $m_2'$. 
We call it  an  error of the second kind.
We now  consider the expected value of
the probability
 of this case, if we use the random encoder  $X^n$
to decode the input message.
We have
\begin{align}&E\left[\mathrm{tr}\left(W_2^{\otimes
n}(X^n(m_1,m_2)){D'}_{X^n(m_1,m_2')}^{(2)}\right)\right]\allowdisplaybreaks\notag\\
&=\mathrm{tr}\left[E\left(W_2^{\otimes
n}(X^n(m_1,m_2))\right)\cdot E\left({D'}_{X^n(m_1,m_2')}^{(2)}\right)\right]\allowdisplaybreaks\notag\\
&= \mathrm{tr}\Bigl[ \rho_2^{\otimes n}
E\left({D'}_{X^n(m_1,m_2')}^{(2)}\right)\Bigr]\allowdisplaybreaks\notag\\
&= \mathrm{tr}\Bigl[ \rho_2^{\otimes n}
 E \left( \Pi_{PW_2, \alpha \sqrt{a}}\Pi_{W_2^{\otimes n}(X^n(m_1,m_2')),\alpha}
\Pi_{PW_2, \alpha \sqrt{a}} \right)\Bigr]\allowdisplaybreaks\notag\\
&= \mathrm{tr}\Bigl[
 E \left( \rho_2^{\otimes n}\Pi_{PW_2, \alpha \sqrt{a}}\Pi_{W_2^{\otimes n}(X^n(m_1,m_2')),\alpha}
\Pi_{PW_2, \alpha \sqrt{a}} \right)\Bigr]\allowdisplaybreaks\notag\\
&= \mathrm{tr}\Bigl[
 E \left((\Pi_{PW_2, \alpha \sqrt{a}} \rho_2^{\otimes n}\Pi_{PW_2, \alpha \sqrt{a}})\Pi_{W_2^{\otimes n}(X^n(m_1,m_2')),\alpha}
 \right)\Bigr]\allowdisplaybreaks\notag\\
&= \mathrm{tr}\Bigl[
(\Pi_{PW_2, \alpha \sqrt{a}} \rho_2^{\otimes n}\Pi_{PW_2, \alpha \sqrt{a}}) E \left(\Pi_{W_2^{\otimes n}(X^n(m_1,m_2')),\alpha}
 \right)\Bigr]\allowdisplaybreaks\notag\\
&\leq 2^{-n[S(\rho_2)-\frac{1}{2}\epsilon]}
\mathrm{tr}\Bigl[\Pi_{PW_2, \alpha \sqrt{a}} E \left(\Pi_{W_2^{\otimes n}(X^n(m_1,m_2')),\alpha}
 \right)\Bigr]\allowdisplaybreaks\notag\\
&\leq 2^{n[\sum_{x\in X}
P(X)S(W_2(\phi_x))-\frac{1}{2}\epsilon]}2^{-n[S(\rho_2)-\frac{1}{2}\epsilon]}
\mathrm{tr}\Bigl[\Pi_{PW_2, \alpha \sqrt{a}}\Bigr]\allowdisplaybreaks\notag\\
&=2^{-n[\sum_{x\in X} P(X)S(W_2(\phi_x))-S(\sum_{x\in X} P(X)W_2(\phi_x))
-\epsilon]}\mathrm{tr}\Bigl[\Pi_{PW_2, \alpha \sqrt{a}}\Bigr]\allowdisplaybreaks\notag\\
&= 2^{-n[\chi(P,\rho^{Y_2})-\epsilon]}\mathrm{tr}\Bigl[\Pi_{PW_2, \alpha \sqrt{a}}\Bigr]\allowdisplaybreaks\notag\\
&\leq 2^{-n[\chi(P,\rho^{Y_2})-\epsilon]}\text{ .}\label{longineq1}
\end{align}
The first equality hold because $X^n(m_1,m_2)$ and  $X^n(m_1,m_2')$
are independent, the first inequality holds because of (\ref{te3}), and
the second inequality holds because of (\ref{te5}). \vspace{0.15cm}

Similarly, let us fix $(m_1',m_2)$, $(m_1,m_2)\in {M'}_1\times {M'}_2$ such
that $m_1\not= m_1'$. Node $1$ would make an error (of the second kind) if $(m_1,m_2)$
has been sent, but node $1$'s  
 decoding results in the message $m_1'$. We now  consider the expected value of the probability
 of this case if we use the random encoder  $X^n$
to decode the input message.
We have
 \begin{equation}E\left[\mathrm{tr}\left(W_1^{\otimes
n}(X^n(m_1,m_2)){D'}_{X^n(m_1,m_2')^{(1)}}\right)\right]
\leq 2^{-n[\chi(P,\rho^{Y_1})-\epsilon]}
 \text{ .}\label{longineq1a}\end{equation}
Thus, the errors of the second kind go to zero if $n$ is sufficiently large.\vspace{0.2cm}

\it Definition of the code:\rm\vspace{0.2cm}

For all $(m_1,m_2)\in {M'}_1\times {M'}_2$ we define
 \[D^{(1)}_{X^n(m_1,m_2)}:=\left(\sqrt{\sum_{m_1^* \in {M'}_1}  {D'}^{(1)}_{X^n(m_1^*,m_2)}}
 \right)^{-1}  {D'}^{(1)}_{X^n(m_1,m_2)} \left(\sqrt{\sum_{m_1^* \in {M'}_1}  {D'}^{(1)}_{X^n(m_1^*,m_2)}} \right)^{-1} \text{ ,}\]
and
\[D^{(2)}_{X^n(m_1,m_2)}:=\left(\sqrt{\sum_{m_2^* \in {M'}_2}  {D'}^{(2)}_{X^n(m_1,m_2^*)}} \right)^{-1}
  {D'}^{(2)}_{X^n(m_1,m_2)} \left(\sqrt{\sum_{m_2^* \in {M'}_2}  {D'}^{(2)}_{X^n(m_1,m_2^*)}} \right)^{-1} \text{ ,}\]
which depends on the random outcome of $X^n$. By  construction,
for any realization $\{x^n(m_1,m_2):m_1\in {M'}_1, m_2\in {M'}_2\}$
of $\{X^n(m_1,m_2):m_1\in {M'}_1, m_2\in {M'}_2\}$ we have for every
$m_1 \in {M'}_1$,
\[\sum_{m_1 \in {M'}_1} D^{(1)}_{x^n(m_1,m_2)} \leq \mathrm{id}_{{H^B}^{\otimes n}}\text{ ,}\] and  for every $m_2 \in {M'}_2$
\[\sum_{m_2 \in M_2} D^{(2)}_{x^n(m_1,m_2)} \leq \mathrm{id}_{{H^C}^{\otimes n}}\text{ .}\]

We combine (\ref{longineq2}) and (\ref{longineq1}), for all $(m_1,m_2)\in {M'}_1\times {M'}_2$ we have
\begin{align}&E\left[\mathrm{tr}\left( D^{(1)}_{X^n(m_1,m_2)} W_1^{\otimes n}(X^n(m_1,m_2))
\right)\right]\allowdisplaybreaks\notag\\
&\geq E\left[\mathrm{tr}\left(  {D'}^{(1)}_{X^n(m_1,m_2)} W_1^{\otimes n}(X^n(m_1,m_2))
\right)\right]\allowdisplaybreaks\notag\\
&- 4 E\left[\mathrm{tr}\left(\sum_{m_1^* \not= m_1}  {D'}^{(1)}_{X^n(m_1^*,m_2)} W_1^{\otimes n}(X^n(m_1,m_2))
\right)\right] \allowdisplaybreaks\notag\\
&\geq 1-\frac{d}{4n\alpha ^2}
-\sqrt{8\frac{ad}{4n\alpha ^2}}\allowdisplaybreaks\notag\\
&- 4 E\left[\mathrm{tr}\left(\sum_{m_1^* \not= m_1}  {D'}^{(1)}_{X^n(m_1^*,m_2)} W_1^{\otimes n}(X^n(m_1,m_2))
\right)\right] \allowdisplaybreaks\notag\\
&\geq 1-\frac{d}{4n\alpha ^2}
-\sqrt{8\frac{ad}{4n\alpha ^2}}-4|{M'}_1| 2^{-n[\chi(P,\sigma^{Y_1})-\epsilon]}\allowdisplaybreaks\notag\\
&\geq 1-\frac{d}{4n\alpha ^2}
-\sqrt{8\frac{ad}{4n\alpha ^2}}-2^{-n\epsilon}\text{ .}\label{cosyt3}
\end{align}
The first inequity holds because of Lemma
\ref{hnopin}.\vspace{0.15cm}

Similarly, if we combine (\ref{longineq2}) and (\ref{longineq1a}), we have  for all
$(m_1,m_2)\in {M'}_1\times {M'}_2$
 \begin{equation}E\left[\mathrm{tr}\left(D^{(2)}_{X^n(m_1,m_2)} W_2^{\otimes n}(X^n(m_1,m_2))
\right)\right]\geq 1-\frac{d}{4n\alpha ^2}
-\sqrt{8\frac{ad}{4n\alpha ^2}}-2^{-n\epsilon}
 \text{ .}\label{cosyt3a}\end{equation}
\vspace{0.2cm}

Since (\ref{cosyt3}) and (\ref{cosyt3a}) hold for all $(m_1,m_2)\in {M'}_1\times {M'}_2$, for any positive $\omega$, choosing a suitable $\alpha$,
if $n$ is sufficiently large,  we have 
\[\sum_{m_1\in {M'}_1}\sum_{m_2\in {M'}_2} \frac{1}{|{M'}_1||{M'}_2|}E\left[\mathrm{tr}\left(D^{(1)}_{X^n(m_1,m_2)} W_1^{\otimes n}(X^n(m_1,m_2))
\right)\right]\geq 1-\omega
 \text{ }\] and
\[\sum_{m_1\in {M'}_1}\sum_{m_2\in {M'}_2} \frac{1}{|{M'}_1||{M'}_2|}E\left[\mathrm{tr}\left(D^{(2)}_{X^n(m_1,m_2)} W_2^{\otimes n}(X^n(m_1,m_2))
\right)\right]\geq 1-\omega
 \text{ .}\]

By the law of large numbers,
if $n$ is sufficiently large, for any positive $\delta$ and $\gamma$, we have
\[p\left\{\sum_{m_1\in {M'}_1}\sum_{m_2\in {M'}_2} \frac{1}{|{M'}_1||{M'}_2|}\mathrm{tr}\left(D^{(1)}_{X^n(m_1,m_2)} W_1^{\otimes n}(X^n(m_1,m_2))
\right)\geq 1-\delta\right\}\geq 1- \gamma\text{ }\] and
\[p\left\{\sum_{m_1\in {M'}_1}\sum_{m_2\in {M'}_2} \frac{1}{|{M'}_1||{M'}_2|}\mathrm{tr}\left(D^{(2)}_{X^n(m_1,m_2)} W_2^{\otimes n}(X^n(m_1,m_2))
\right)\geq 1-\delta\right\}\geq 1- \gamma\text{ .}\] Thus,
\begin{align*}&p\Bigl\{\sum_{m_1\in {M'}_1}\sum_{m_2\in {M'}_2} \frac{1}{|{M'}_1||{M'}_2|}\mathrm{tr}\left(D^{(1)}_{X^n(m_1,m_2)} W_1^{\otimes n}(X^n(m_1,m_2))
\right)\geq 1-\delta \text{ and }\\
&\sum_{m_1\in {M'}_1}\sum_{m_2\in {M'}_2} \frac{1}{|{M'}_1||{M'}_2|}\mathrm{tr}\left(D^{(2)}_{X^n(m_1,m_2)} W_2^{\otimes n}(X^n(m_1,m_2))
\right)\geq 1-\delta\Bigr\}\\
&\geq 1- 2\gamma \text{ .}\end{align*}

If $n$ is sufficiently large, with a positive probability, we can
find a realization $x^n(m_1,m_2)$ of $X^n(m_1,m_2)$  such that
\[\sum_{m_1\in {M'}_1}\sum_{m_2\in {M'}_2} \frac{1}{|{M'}_1||{M'}_2|}\mathrm{tr}\left(D^{(1)}_{x^n(m_1,m_2)} W_1^{\otimes n}(x^n(m_1,m_2))
\right)\geq 1-\delta\text{ ,}\] and
\[\sum_{m_1\in {M'}_1}\sum_{m_2\in {M'}_2} \frac{1}{|{M'}_1||{M'}_2|}\mathrm{tr}\left(D^{(2)}_{x^n(m_1,m_2)} W_2^{\otimes n}(x^n(m_1,m_2))
\right)\geq 1-\delta\text{ .}\]\vspace{0.2cm}

\it Definition of the message sets:\rm\vspace{0.2cm}

Assume 
\begin{align*}&\biggl\lvert \Bigl\{ m_2\in {M'}_2 : \sum_{m_2\in {M'}_2} \frac{1}{|{M'}_2|}\mathrm{tr}\left(D^{(1)}_{x^n(m_1,m_2)} W_1^{\otimes n}(x^n(m_1,m_2))
\right)< 1-2\delta
\Bigr\}\biggr\rvert\\
&> \frac{1}{2} |{M'}_2|\text{ .}\end{align*} 
We  have in this case,
\[\sum_{m_1\in {M'}_1}\sum_{m_2\in {M'}_2} \frac{1}{|{M'}_1||{M'}_2|}\mathrm{tr}\left(D^{(1)}_{x^n(m_1,m_2)} W_1^{\otimes n}(x^n(m_1,m_2))
\right)< 1-\delta\text{ ,}\] but this is a contradiction to the result above.

Thus, there exists a set $M_2 \in {M'}_2$ such that $|M_2| = \lceil
\frac{1}{2} |{M'}_2|\rceil$ and for every $m_2 \in M_2$ we have

\begin{equation} \sum_{m_1\in {M'}_1} \frac{1}{|{M'}_1|}\mathrm{tr}\left(D^{(1)}_{x^n(m_1,m_2)} W_1^{\otimes n}(x^n(m_1,m_2))
\right)\geq 1-2\delta\text{ .}\label{ineuqfour1}\end{equation}

Similarly,  there exists a set $M_1 \in {M'}_1$ such that $|M_1| =
\lceil \frac{1}{2} |{M'}_1|\rceil$ and for every $m_1 \in M_1$ we
have
\begin{equation} \sum_{m_2\in {M'}_2} \frac{1}{ |{M'}_2|}\mathrm{tr}\left(D^{(2)}_{x^n(m_1,m_2)} W_2^{\otimes n}(x^n(m_1,m_2))
\right)\geq 1-2\delta\text{ .}\label{ineuqfour2}\end{equation}\vspace{0.15cm}



For every $(m_1,m_2)\in M_1\times M_2$, we define
\begin{equation}w((m_1,m_2)):=x^n(m_1,m_2)\text{ ,}\end{equation}
\begin{equation}D^{(m_1)}_{m_2} :=
D^{(1)}_{x^n(m_1,m_2)}\text{ ,}\end{equation} and
\begin{equation}
D^{(m_2)}_{m_1} := D^{(2)}_{x^n(m_1,m_2)}\text{
.}\end{equation}

$\{D^{(m_2)}_{m_1}:m_1 \in M_1\}$ is less or equal to
 the partition of the identity for every $m_2\in M_2$. 
$\{D^{(m_1)}_{m_2}:m_2 \in M_2\}$ is less or equal to
 the partition of the identity for every $m_1\in M_1$.
\vspace{0.15cm}

Since node $1$ already knows the message $m_2\in M_2$, it  chooses
the corresponding decoding set
\[\left\{D^{(m_2)}_{m}:m \in M_1\right\}\] to decode $m_1 \in M_1$.
Since node $2$ already knows the message $m_1\in M_1$, it  chooses
the corresponding decoding set
\[\left\{D^{(m_1)}_{m}:m \in M_2\right\}\] to decode $m_2 \in M_2$.

By (\ref{ineuqfour1}) and (\ref{ineuqfour2}),
for every $m_2 \in M_2$ we have
\begin{equation}\sum_{m_1\in {M}_1} \frac{1}{|{M}_1|}\mathrm{tr}\left(D^{(m_2)}_{m_1} W_1^{\otimes n}(x^n(m_1,m_2))
\right)\geq 1-4\delta\text{ ,}\end{equation} and for every $m_1 \in M_1$ we have
\begin{equation} \sum_{m_2\in {M}_2} \frac{1}{ |{M}_2|}\mathrm{tr}\left(D^{(m_1)}_{m_2} W_2^{\otimes n}(x^n(m_1,m_2))
\right)\geq 1-4\delta\text{ .}\end{equation}
Thus, for all sufficiently large $n\in\mathbb{N}$ any rate pairs  satisfying
\[R_1 \leq \chi(P;\sigma^{Y_1})  - 2\epsilon -\frac{1}{n}  \text{ }\] and
\[ R_2 \leq   \chi(P;\sigma^{Y_2})  - 2\epsilon -\frac{1}{n}\text{ }\]
are achievable.\qed\vspace{0.2cm}

If we combine Lemma \ref{macphase} and Theorem \ref{broadphase}, we
obtain

\begin{corollary}
Let $N$ be a two-phase bidirectional relaying quantum channel. Let $H^{Y_1}$ be the
Hilbert space whose unit vectors correspond to the pure states of node $1$'s quantum
system,  $H^{Y_2}$ be the
Hilbert space whose unit vectors correspond to the pure states of node $2$'s quantum
system, and $H^{X}$ be the
Hilbert space whose unit vectors correspond to the pure states of the relay node's quantum
system.

We assume that the relay node's encoding is restricted to transmitting an
indexed finite set of orthogonal quantum  states $X \subset H^X$.

We assume that  node $1$'s encoding is restricted to transmitting an indexed
finite set of orthogonal quantum  states $Y_1\subset H^{Y_1}$.

We assume that node $2$'s encoding is restricted to transmitting an indexed
finite set of orthogonal quantum  states $Y_2\subset H^{Y_2}$.

The classical-quantum capacity region  of the  two-phase bidirectional relaying quantum channel $N$
 with average error is the
 intersection of  two rate regions,  Region $1$ and  Region $2$, which
are defined as follows:\\[0.15cm]
\bf 1\rm:   Region $1$ is the set of all rate pairs $(R_1, R_2)$ such that
\begin{equation} R_2 \leq  \chi(Q_1;\sigma^{X}) \text{ ,}
\label{Brate2}\end{equation}
\begin{equation} R_1 \leq \chi(Q_2;\sigma^{X}) \text{ ,}
\label{Brate1}\end{equation} and
\begin{equation} R_2 + R_1 \leq  \chi(Q_{1,2};\sigma^{X})\label{Brate3}\end{equation}
for any joint probability distribution $Q_{1,2}$ on $Y_1\times Y_2$.
Here $Q_1$ is the marginal probability distribution of $Q_{1,2}$ on $Y_1$,
 $Q_2$ is the marginal probability distribution of $Q_{1,2}$ on $Y_2$, and
$\sigma^{X}$ is the resulting quantum state at the outcome of the relay node.\\[0.15cm]
\bf  2\rm:  Region $2$ is the set of all rate pairs $(R_1, R_2)$ such that
\begin{equation} R_1 \leq  \limsup_{n\rightarrow \infty} \frac{1}{n}\chi(P^n;{\sigma^{Y_1}}^{\otimes n})    \text{ }
\label{Brate4}\end{equation} and
\begin{equation} R_2 \leq  \limsup_{n\rightarrow \infty} \frac{1}{n}\chi(P^n;{\sigma^{Y_2}}^{\otimes n})  \text{ }
\label{Brate5}\end{equation}
for all probability
distribution $P$ on $X$.
Here, $\sigma^{Y_1}$ is the resulting quantum state at the outcome of node $1$,
while $\sigma^{Y_2}$ is the resulting quantum state at the outcome of node $2$.
\label{bidirec}\end{corollary}

\begin{remark}
Note  that the capacity region
 of a multiple-access channel with
 average errors is not equal to its capacity region
with maximal errors. 
This is a  well-known fact in the 
 classical information theory (cf. \cite{Due} and \cite{Ahl5}).
We consider  average errors, not maximal errors in  Theorem
\ref{broadphase} and Corollary \ref{bidirec}, since we use Lemma
\ref{macphase}, which
 considered average errors, for the
multiple-access phase. 
\label{avemax}
\end{remark}

\begin{remark}Without loss of generality, 
we assume that $\chi(P;{\sigma^{Y_1}})  \geq {\chi(P;\sigma^{Y_2}})$,
i.e., $W_2$, the channel which connects  the relay node and node $2$, 
has a lower capacity than $W_1$ in the broadcast phase.
If  $\chi(Q_2;\sigma^{X})= {\chi(P;\sigma^{Y_2}})$, 
i.e., the capacities of $W_2$ in both directions are identical, then
$R_1$ cannot exceed $\chi(P;{\sigma^{Y_2}})$ in the multiple-access phase.
In this case, we may assume that in the broadcast phase the message
sets $M_1=\{1,\ldots,|M_1|\}$ and $M_2=\{1,\ldots,|M_2|\}$, that  the relay node sends to node $1$
and node $2$, satisfy $|M_1| \leq
2^{n\chi(P;{\sigma^{Y_2}})-\epsilon}$ and $|M_2|\leq
2^{n\chi(P;{\sigma^{Y_2}})-\epsilon}$ for a positive $\epsilon$.

In this case, we have a very simple coding strategy for the
broadcast phase. The common message set which the relay node sends to
both node $1$ and node $2$ in the broadcast phase is a set $M'=\{1,\ldots,|M'|\}$
which satisfies $|M'| = \lfloor2^{n\chi(P;\sigma^{Y_2})-\epsilon}\rfloor$.

We consider the case that the relay node wants to send $(m_1,m_2)\in
M_1\times M_2$, where node $1$ shall detect $m_1$, while  node $2$
shall detect $m_2$.
 Then, the relay
node sends $m_1+m_2 \mod |M'|$ as a common message to both node $1$
and node $2$. By the HSW Random Coding Theorem (cf. \cite{Sch/We}
and \cite{Ho2}) node $1$ and node $2$ can decode the common message
if the size of the message set is less than
$2^{n\chi(P;\sigma^{Y_2})}$ 

Since node $1$ already knows $m_2$, it can obtain $m_1$ by simply
subtracting  $m_2$  from $m_1+m_2$ modulo $|M'|$. Since node $2$
already knows $m_1$, it can obtain $m_2$ by subtracting  $m_1$ from
$m_1+m_2$ modulo $|M'|$.
\label{remnet}
\end{remark}

\section*{Acknowledgment}
Support by the Bundesministerium f\"ur Bildung und Forschung (BMBF)
via Grant 16BQ1050 and  16BQ1052 is gratefully acknowledged.

\end{document}